\documentclass[prb,twocolumn,preprintnumbers,amsmath,amssymb,superscriptaddress]{revtex4-1}

\usepackage{epsf}
\usepackage{graphicx}
\usepackage{bm}
\usepackage{amsmath}
\usepackage{color}
\usepackage{notes2bib}
\usepackage{epstopdf}
\usepackage{textcomp}

\usepackage{physics}
\usepackage{braket}

\DeclareMathOperator{\sgn}{sign}

\begin{document}

\title{Chiral spin structure of electron gas in systems with magnetic skyrmions} 
\author{L.~A.~Yung}
\email{levyung@gmail.com}
\affiliation{Ioffe Institute, 194021 St.Petersburg, Russia}
\author{K.~S.~Denisov}
\email{denisokonstantin@gmail.com} \affiliation{Ioffe Institute, 194021 St.Petersburg, Russia}
\affiliation{Lappeenranta-Lahti University of Technology, FI-53851 Lappeenranta, Finland}
\author{I.~V.~Rozhansky}
\affiliation{Ioffe Institute, 194021 St.Petersburg, Russia}
\affiliation{Lappeenranta-Lahti University of Technology, FI-53851 Lappeenranta, Finland}
\author{N.~S.~Averkiev}
\affiliation{Ioffe Institute, 194021 St.Petersburg, Russia}
\author{E.~L\"ahderanta}
\affiliation{Lappeenranta-Lahti University of Technology, FI-53851 Lappeenranta, Finland}

\begin{abstract}
The theoretical study considers chiral spin texture induced in a 2D electron gas (2DEG)
by  magnetic skyrmions. 
We calculate the electron gas spin density as a linear response to the exchange interaction between the 2DEG and the magnetization field of a magnetic skyrmion.
Two physically distinct regimes occur. 
When the size of the skyrmion is larger than the inverse Fermi wavevector $k_F^{-1}$, the spin density response follows the magnetization profile of the skyrmion. 
In the opposite case of a small skyrmion the emerging spin structure of 2DEG 
has a characteristic size of $k_F^{-1}$ and the response becomes non-local, it can be 
viewed as chiral Friedel oscillations. At that, the emerging spin structure of the
oscillations appears to be more complex than that of the skyrmion itself.
\end{abstract}

\maketitle
\section{Introduction}
Magnetic skyrmions are stable chiral spin excitations which are formed in ferromagnetic thin films with a significant Dzyaloshinki-Morya interaction. 
Nowadays skyrmions are  intensively studied in various systems~\cite{NagaosaNature,fert2013skyrmions,fert2017magnetic,wiesendanger2016nanoscale,Uzdin-1,Wiesendanger-1,rozsa2016complex}.
Along with the fundamental interest 
magnetic skyrmions are recognized to have a potential in practical 
applications in magnetic memory. 
Skyrmions are topologically protected  spin textures, hence, they are rather robust.
Having also small size, which can be decreased at best below ten nanometers~\cite{Weisendanger2018}
they are attractive for storage applications such as magnetic racetrack memory~\cite{fert2017magnetic}.
Skyrmions along with other chiral configurations of immobile magnetic moments in a ferromagnet have been widely studied both theoretically and experimentally with a focus on their fabrication, observation and dynamics~\cite{fert2017magnetic,wiesendanger2016nanoscale,soumyanarayanan2017tunable,nakajima2017skyrmion}
as well as the electron transport properties of skyrmionic system including topological Hall effect~\cite{BrunoDugaev,denisov2018general,QTHE}.

Much less attention has been focused on the equilibrium properties of the electron gas itself due to an exchange coupling with chiral spin textures. 
It has been realized  that chiral electron gas oscillations would appear in a vicinity of 
a defect or an impurity in a solid with broken time reversal symmetry~\cite{ourSci2019}.
It has been shown in Ref.~\cite{ourSci2019} that a 2D electron gas with a parabolic or Dirac-like 
energy spectra with spin-orbit and exchange spin splitting forms chiral Friedel-like oscillations of spin density around 
a charged defect. 
One can expect a similar effect for a free electron gas interacting with magnetic skyrmions,  
having also a prominent impact on the whole system. 
For instance, it has been recently demonstrated that the redistribution of an 
electron gas spin density in a multilayered system can lead to a specific RKKY interaction between skyrmions in  neighboring layers~\cite{Cacilhas2018}. 
The study of general features associated with a free carriers spin polarization 
pattern driven by magnetic skyrmions is thus of a high interest.
It is particularly timely in light of the recent 
advantages in spin polarized scanning tunneling microscopy with  
its state of the art capability to resolve  
chiral spin textures below 10 nm in size~\cite{Weisendanger2019}. An experiment aiming at observation of
the real space structure of an electron spin density 
induced by chiral spin textures therefore would
give a direct access to the parameters of the electron gas and magnetic properties in such a system.    


In this work we show that the chiral oscillations of the electron spin density exist for a purely parabolic electron spectra with no spin-orbit interaction if the defect itself is chiral. 
We consider the exchange interaction  of free electrons in a 2D  ferromagnetic film with skyrmions. Using the linear response theory we show that a magnetic skyrmion induces chiral Friedel-like oscillations of the free electron spin density. At that there can be two distinct cases. For a large skyrmion the response
   of the electrons spin density is local, it follows the magnetization profile of the skyrmion. 
   A different situation appears when the skyrmion size $a$ goes below the inverse Fermi wavevector $k_F^{-1}$. At that the response becomes non-local and the spatial size of the chiral perturbation 
   of the electron spin density maintains the characteristic size of $k_F^{-1}$. 

   

\section{Model}
Let us consider a degenerate 2DEG  coupled by exchange interaction with a skyrmion in a ferromagnetic layer.
The Hamiltonian of 2DEG reads:
\begin{equation}
    \hat{H} = \frac{\hat{\boldsymbol{p}}^2}{2m} - \alpha \bar{\boldsymbol{M}}(\boldsymbol{r}) \cdot \hat{\boldsymbol{\sigma}},
\end{equation}
where 
$\hat{\bm{p}}$ is the momentum operator, $m$ is the electron effective mass,
$\alpha$ is the exchange interaction constant, $\bar{\boldsymbol{M}}(\boldsymbol{r})$ is the skyrmion 
field and $\hat{\boldsymbol{\sigma}}$ is the vector of Pauli matrices acting on electron spin. 
The skyrmion 
field $\bar{\boldsymbol{M}}(\boldsymbol{r})$ has the following form
\begin{equation} \label{eq:2}
\bar{\boldsymbol{M}}(\boldsymbol{r}) = \left(\begin{matrix}
\bar{M}_{\parallel}(r) \cos (\kappa \varphi + \gamma)\\
\bar{M}_{\parallel}(r) \sin (\kappa \varphi + \gamma)\\
\bar{M}_{z}(r)
\end{matrix}\right),\,\,\,\left\{\begin{matrix}\bar{M}_{\parallel}(r) = \sin \theta(r)\\
\bar{M}_{z} (r) = \cos \theta(r)
\end{matrix}\right.,
\end{equation}
where $r$ and $\varphi$ are the polar coordinates in the film plane $\bm{r}=(r,\varphi)$,  $\kappa \in \left\{0, \pm 1, \pm 2, ...\right\}$  is the skyrmion vorticity, $\gamma \in \{0...2\pi\}$ is the
skyrmion helicity. 
$\theta (r)$ is the out of plane angle of the magnetization field; the
following boundary conditions are assumed: $\left.\theta (r)\right|_{r = 0} =\pi $ and $ \left.\theta (r)\right|_{r = \infty} = 0$, which by the definition of $\bar{\boldsymbol{M}}$ result in
\begin{equation}\left.\bar{\boldsymbol{M}}(\boldsymbol{r})\right|_{r = 0} = - \boldsymbol{e}_z \text{,   } \left.\bar{\boldsymbol{M}}(\boldsymbol{r})\right|_{r = \infty} = \boldsymbol{e}_z.
\end{equation}
$\boldsymbol{e}_z$ is the unit vector in out of plane $z$ direction.

\begin{figure}[t]
    \centering
    \includegraphics[scale = 0.5]{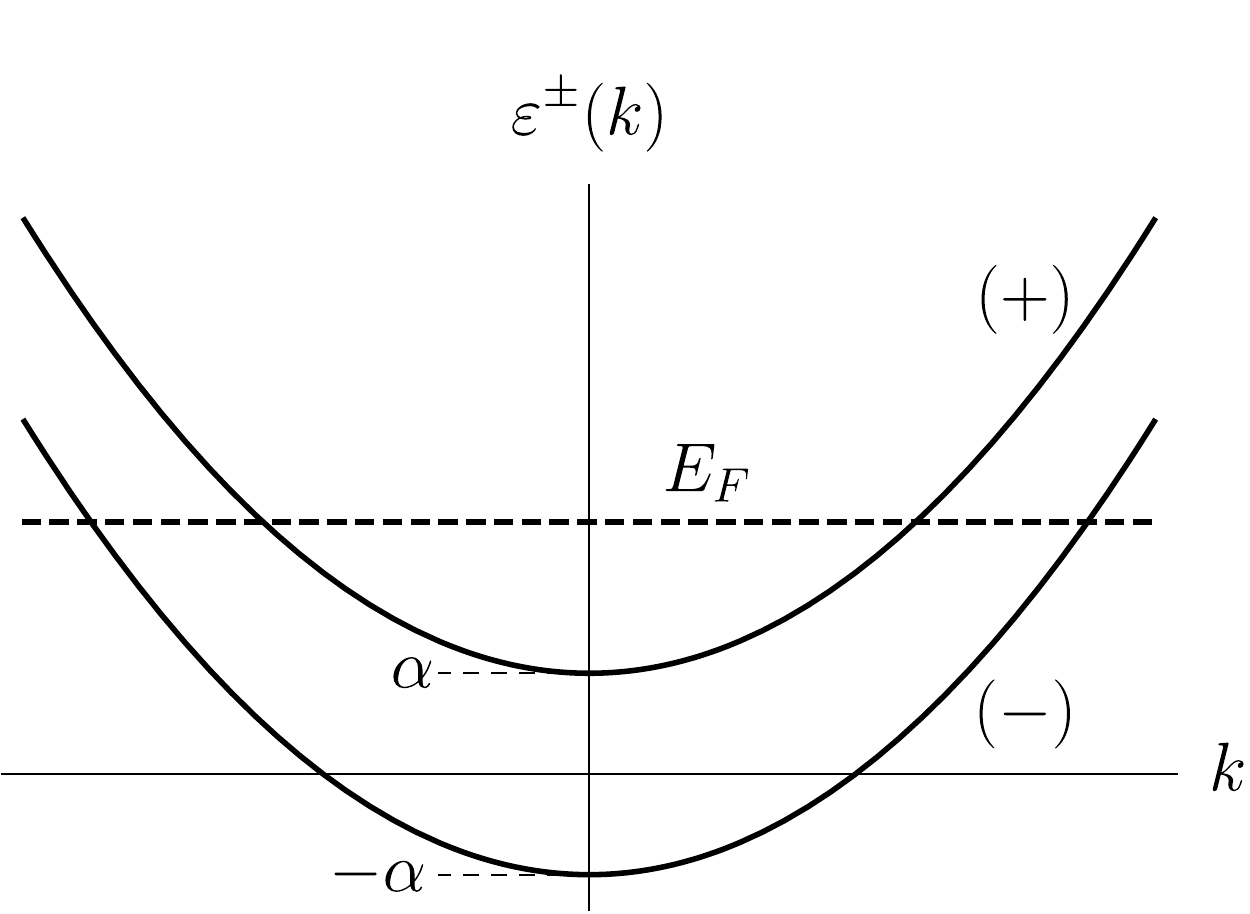}
    \caption{Parabolic spectrum with two subbands of spin-up (-) and spin down (+)  states.}
    \label{fig:1}
\end{figure}

We treat the electron gas exchange interaction with a skyrmion core as a perturbation $\hat{V}$, 
at that the unperturbed Hamiltonian $\hat{H}_0$ of 2DEG 
includes spin splitting due to the uniform background magnetization:
\begin{gather}\label{eq:5}
\hat{H} = \hat{H}_0 + \hat{V},     \\
\hat{H}_0 = \frac{\hat{\boldsymbol{p}}^2}{2m} - \alpha \hat{\sigma}_z,
\quad 
\hat{V} = - \alpha \boldsymbol{M}(\boldsymbol{r}) \cdot \hat{\boldsymbol{\sigma}},
\notag
\end{gather}
where $\boldsymbol{M}(\boldsymbol{r}) \equiv \bar{\boldsymbol{M}}(\boldsymbol{r})-\boldsymbol{e}_z$. 
The Hamiltonian $\hat{H}_0$ has parabolic spectrum with two shifted spin subbands:
(see Fig.~\ref{fig:1}):
\begin{gather}
\varepsilon_{\boldsymbol{p}}^{\pm} = \frac{p^2}{2 m} \pm \alpha,
\end{gather}
here index $\pm$ corresponds to $\ket{\downarrow}$, $\ket{\uparrow}$ states.
With $E_F$ being the Fermi energy the Fermi wavevectors for each spin subband are given by:
\begin{gather}
k_F^\pm = k_F \sqrt{ 1 \mp \alpha/E_F},
\quad 
k_F = \sqrt{\frac{2m E_F}{\hbar^2}}.
\end{gather}

In this paper we analyze
the structure of 2DEG spin density $\delta \boldsymbol{S}(\boldsymbol{r})$ emerging around a magnetic skyrmion   $\boldsymbol{M}(\boldsymbol{r})$. 
As a starting point we make use of the linear coupling between $\delta \boldsymbol{S}(\boldsymbol{q}), \boldsymbol{M}(\boldsymbol{q})$ 
Fourier images:

\begin{equation} \label{eq:11}
\delta \boldsymbol{S}(\boldsymbol{q}) = \alpha \hat{\chi}(\boldsymbol{q}) \boldsymbol{M}(\boldsymbol{q}),
\end{equation} 
where 
$\hat{\chi}$ is a spin susceptibility matrix, which depends on the electron distribution function and the $\hat{H}_0$ spectrum. 


\section{Results and Discussion}
The spin density response in real space can be found from:
\begin{equation} \label{eq:12}
\delta \boldsymbol{S}(\boldsymbol{r}) =\alpha \int \frac{d\boldsymbol{q}}{2\pi} e^{i \boldsymbol{q} \boldsymbol{r}} \hat{\chi}(\boldsymbol{q}) \boldsymbol{M}(\boldsymbol{q}).
\end{equation} 
 Note that $\hat{\chi}$ depends only on $q=|\boldsymbol{q}|$ (the 2DEG is homogeneous), 
therefore  $\delta\boldsymbol{S}(\boldsymbol{r})$ can be written in the form
\begin{equation} \label{eq:13}
\delta{\boldsymbol{S}}(\boldsymbol{r}) = \left(\begin{matrix}
{\delta S}_{\parallel}(r) \cos (\kappa \varphi + \gamma)\\
{\delta S}_{\parallel}(r) \sin (\kappa \varphi + \gamma)\\
{\delta S}_{z}(r)
\end{matrix}\right).
\end{equation}
With this being said the equations (\ref{eq:11}) and (\ref{eq:12}) can be rewritten as
\begin{equation}  
\delta {S_{\parallel, z}}(q) = \alpha \hat{\chi}(q) M_{\parallel, z}(q),
\end{equation} 
\begin{equation} 
\delta S_{\parallel, z}(r) =\alpha \int \frac{d q}{2\pi} J_{1,0}(qr) \hat{\chi}(q) M_{\parallel, z}(q).
\end{equation} 
Let us note that the direction of the in-plane component of the electron spin density $\delta \boldsymbol{S}_{x,y}(\boldsymbol{r})$ 
coincides with that of the skyrmion magnetization. This is a consequence of the exchange coupling, 
whereas spin-orbital coupling would have resulted  an additional phase shift added to $\gamma$. 
However, the spin texture formed by the electron spin density does not necessarily follow the skyrmion 
structure. The additional phase shift emerges in the case when the coupling becomes non-local. 

The behaviour of $\delta \bm{S}_{z,\parallel}({r})$ in real space strongly depends 
both on $k_F a$ (here $a$ the skyrmion radius), which defines a spatial scale, and on 
${\alpha}/{E_F}$ which controls the filling of the spin split subbands (Fig.~\ref{fig:1}).
In order to analyze these dependencies 
we calculate numerically $\delta \bm{S}_{z,\parallel}({r})$ using 
the skyrmion radial profile $\theta(r)$ (Ref.~\cite{Romming,bessarab2018lifetime})
\begin{equation}
\theta (r) = -2 \arcsin\left( \tanh \frac{r}{a/2}\right) + \pi,
\end{equation}
where $a > 0$ is the radius of $\boldsymbol{M}(\boldsymbol{r})$ localization.


\begin{figure*}[t]
    \centering
    \includegraphics[scale = 0.6]{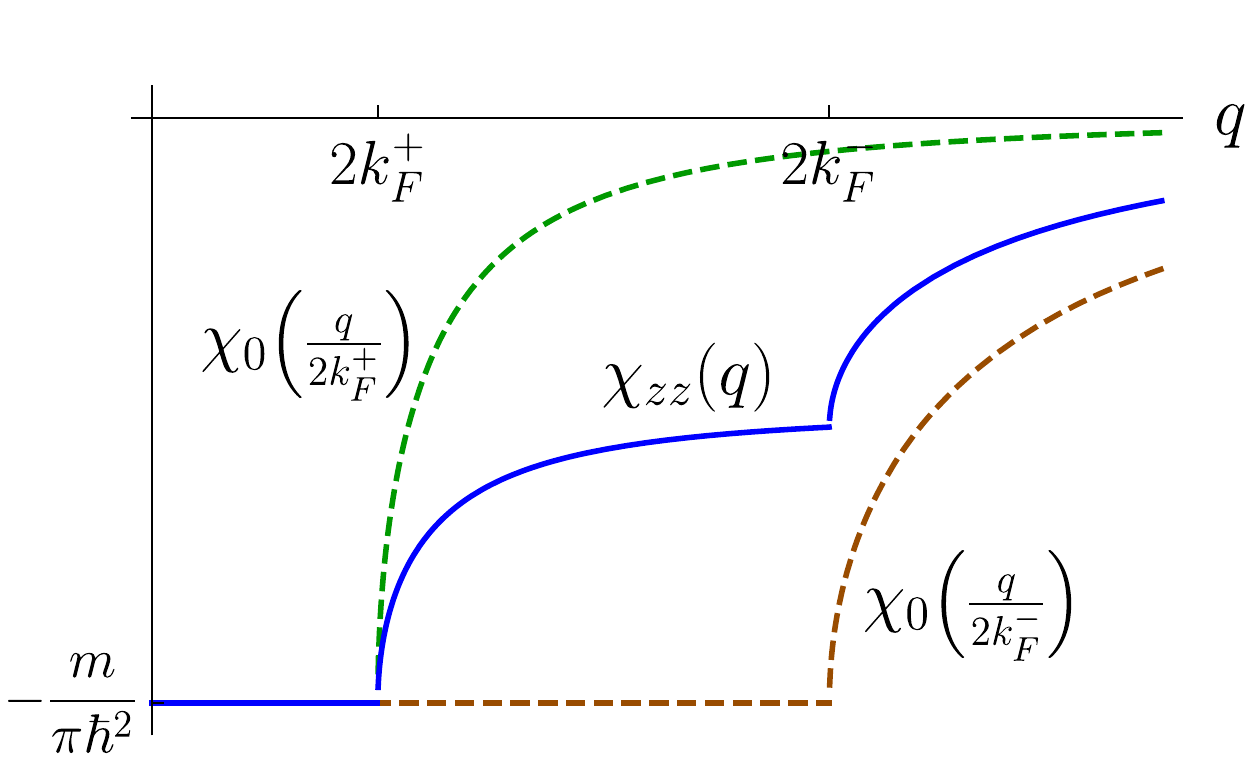}
    \includegraphics[scale = 0.6]{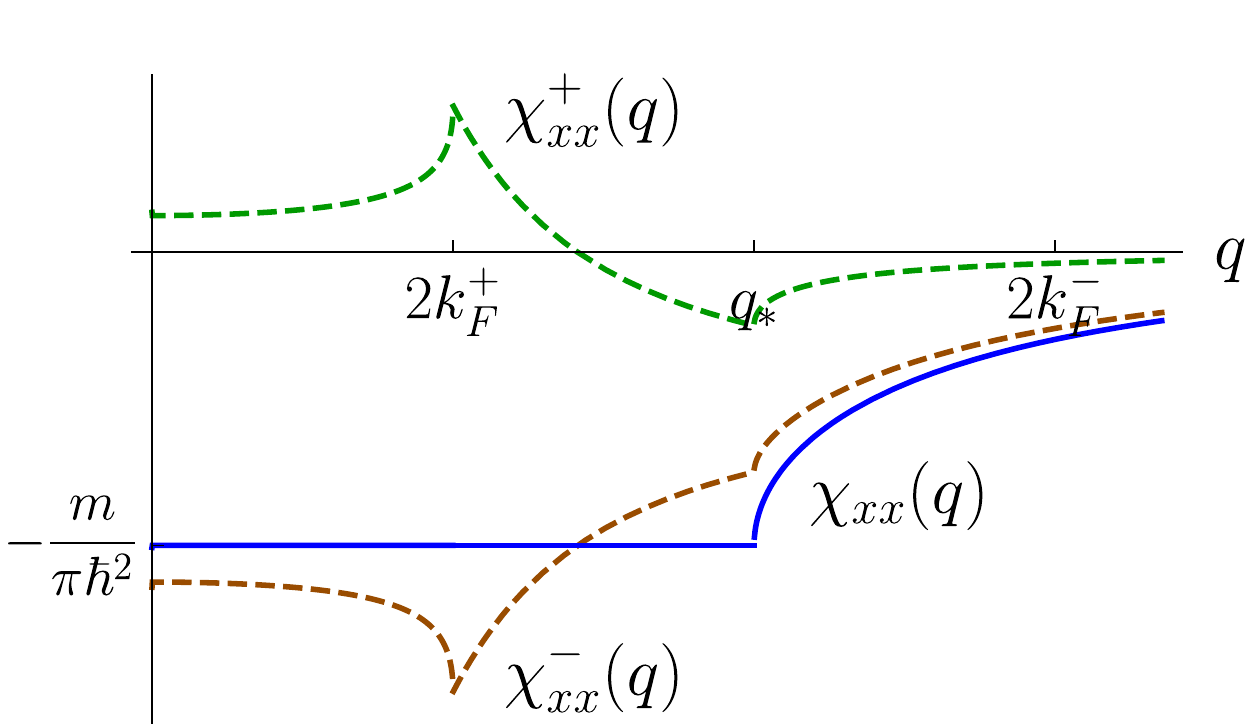}
    \caption{Susceptibility functions $\chi_{z, \parallel}$  at ${\alpha}/{E_F} = 0.8$ (two subbands are filled).}
    \label{fig:2}
\end{figure*}

\subsection{Two subbands}
Let us  consider the case when the Fermi energy $E_F > \alpha$, so that 
both spin subbands are occupied. The spin susceptibility matrix has only diagonal non-zero elements
\begin{equation}
    \hat{\chi}(q) = \begin{pmatrix}
    \chi_\parallel(q) & 0 & 0
    \\
0 & \chi_\parallel(q) & 0
    \\
    0& 0 & \chi_z(q) 
    \end{pmatrix},
\end{equation} 
The functions $\chi_{z,\parallel}(q)$ are given by (the details of derivation are summarized in the Appendix A):
\begin{equation} \label{eq:18}
\chi_\parallel = \chi_{\parallel}^+ \left(\frac{q}{2 k_F^+}\right) + \chi_{\parallel}^- \left(\frac{q}{2 k_F^-}\right),
\end{equation}
\begin{equation} \label{eq:19}
\chi_{z} = \frac{1}{2} \left(\chi_0\left(\frac{q}{2 k_{F}^+}\right) + \chi_0\left(\frac{q}{2 k_{F}^-}\right)\right),
\end{equation}
where the spin susceptibility functions for each of the subbands are as follows
\begin{multline}
\chi_\parallel^\pm (z) = -\frac{m}{2 \hbar^2 \pi} \frac{1}{z} \left[\left(z \mp \frac{\alpha_\pm}{z}\right) - \theta\left(\left|z \mp \frac{\alpha_\pm}{z}\right| - 1\right) \times\right.\\\left. \times\sqrt{\left(z \mp \frac{\alpha_\pm}{z}\right)^2 -1} \cdot \sgn\left(z \mp \frac{\alpha_\pm}{z}\right)\right],
\end{multline}
\begin{equation}
\chi_0 (z) = -\frac{m}{\hbar^2 \pi} \left[1 - \theta(z - 1) \sqrt{1 - 1/z^2}\right].
\end{equation}
Here  $\alpha_{\pm} = {\alpha}/{2(E_F \mp \alpha)}$, 
the function $\chi_0\left({q/}{2 k_{F}}\right)$ is an ordinary 2D Lindhard function, 

\begin{figure*}[t]
    \centering
    \includegraphics[scale = 0.6]{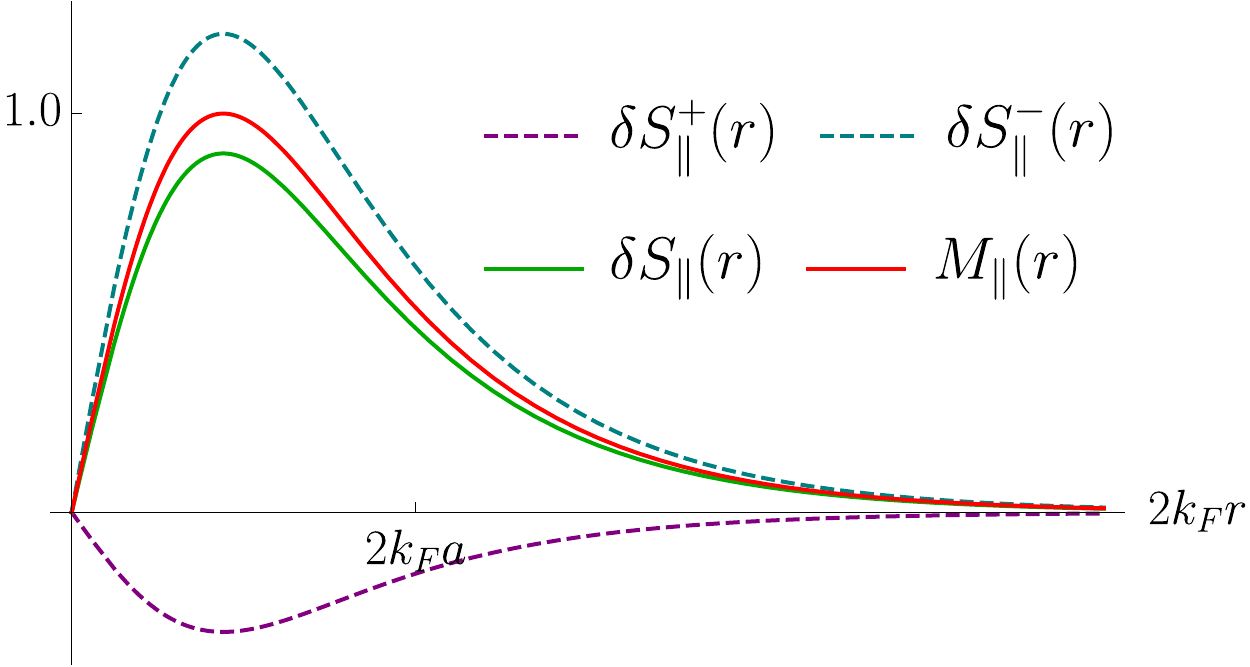}
    \includegraphics[scale = 0.6]{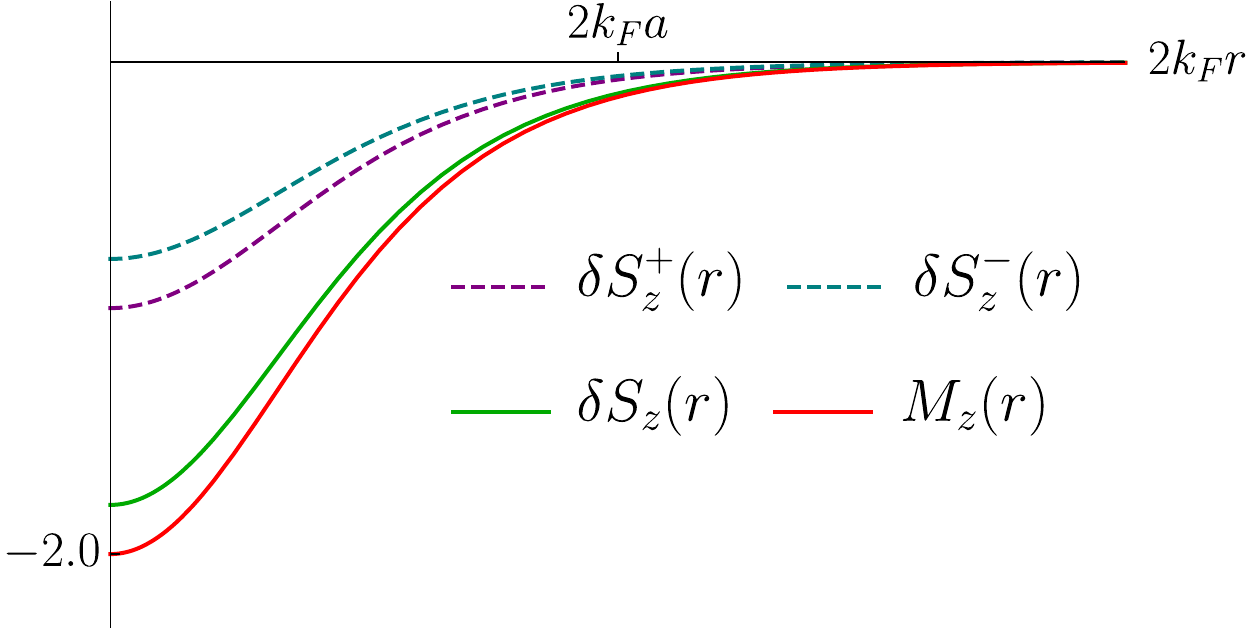}
    \caption{Spin density response $\delta S_{\parallel,z}$  induced by a large skyrmion: $k_F a = 10$. Two subbands are filled: $\alpha/E_F = 0.8$.}
    \label{fig:3}
\end{figure*}


Let us comment on the structure of the out-of-plane and in-plane spin susceptibilities $\chi_{z,\parallel}$.
The function $\chi_{z}$  is a half-sum of two $\chi_0$ Lindhard functions with different $k^{\pm}_F$ wave-vectors in the argument (see Fig. \ref{fig:3}). As a result $\chi_{z}$ has two critical points corresponding to the nesting vectors of the Fermi surface,
so the spin density $\delta S_z$ would exhibit Friedel oscillations with two spatial periods given by $\pi/k_F^{-}$ and $\pi/k_F^{+}$.
The function $\chi_{\parallel}$  has a more complex structure.
Although both spin subbands are filled, the total $\chi_{\parallel}$
function turns out  to have 
only one critical point, thus, only one type of oscillations is expected, the corresponding period is  $2\pi/q_*$, where $q_\ast$ is given by:
\begin{gather}
q_* = 2k_F \cdot \sqrt{\frac{1 + \sqrt{1- 4 \beta^2}}{2}},\,\,\,\,\,
2k_F^+ \leq q_* \leq  2k_F^-.
\end{gather}

The susceptibilities introduced above allow us 
to analyze the spatial structure of 2DEG spin density nearby a magnetic skyrmion.
Two regimes can be distinguished depending on $k_F a$ parameter.
If $k_F a \gg 1$ (equivalent to $k_F^\pm a \gg 1$, as ${\alpha}/{E_F} < 1$) the  $M_{\parallel,z}(q)$ in the momentum representation is localized within the range 
$q\ll k_F$.   
Figures (\ref{fig:2}) show that the spin susceptibility functions are constants within the scale of $k_F^+$. 
Consequently, in this case the formula (Eq. \ref{eq:11}) shows that $\delta \boldsymbol{S} (\boldsymbol{r})$ and $\sim \boldsymbol{M} (\boldsymbol{r})$ are locally coupled. 
Fig.~\ref{fig:3}  shows the calculated spin density response  $\delta \boldsymbol{S} (\boldsymbol{r})$ following 
 the skyrmion magnetization profile 
$\boldsymbol{M} (\boldsymbol{r})$. The amplitude of Friedel oscillations in this regime is negligible compared to the amplitude of the perturbed electron spin density. 

\begin{figure*}[t]
    \centering
    \includegraphics[scale = 0.6]{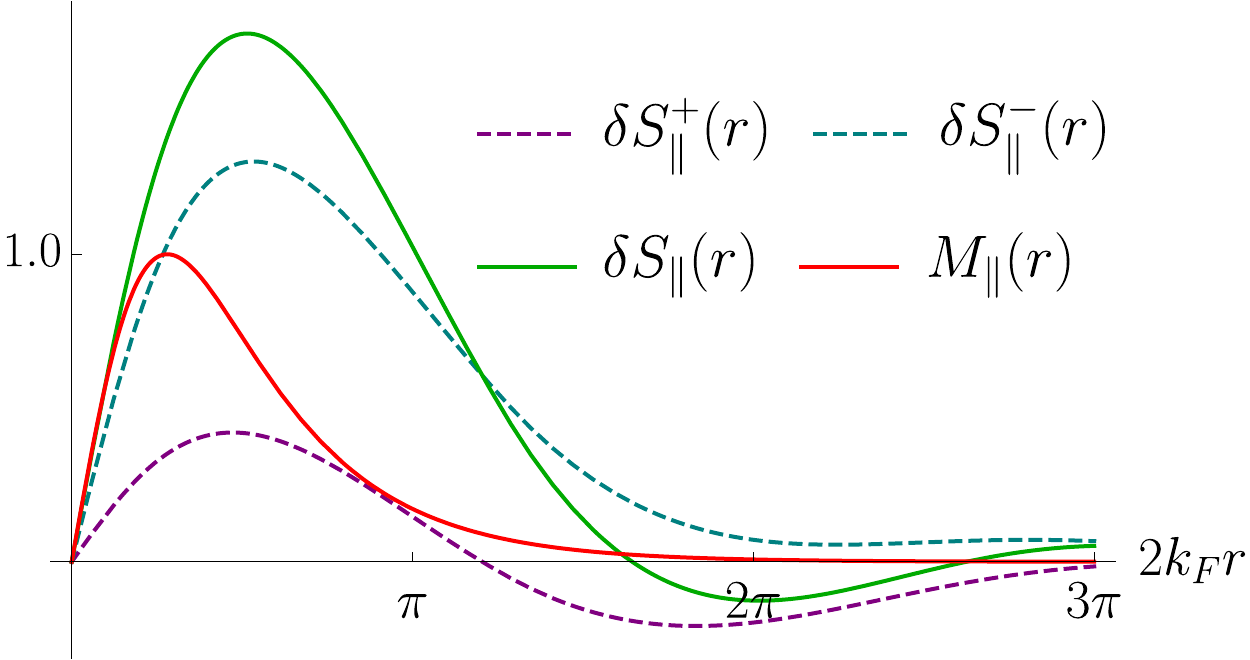}
    \includegraphics[scale = 0.6]{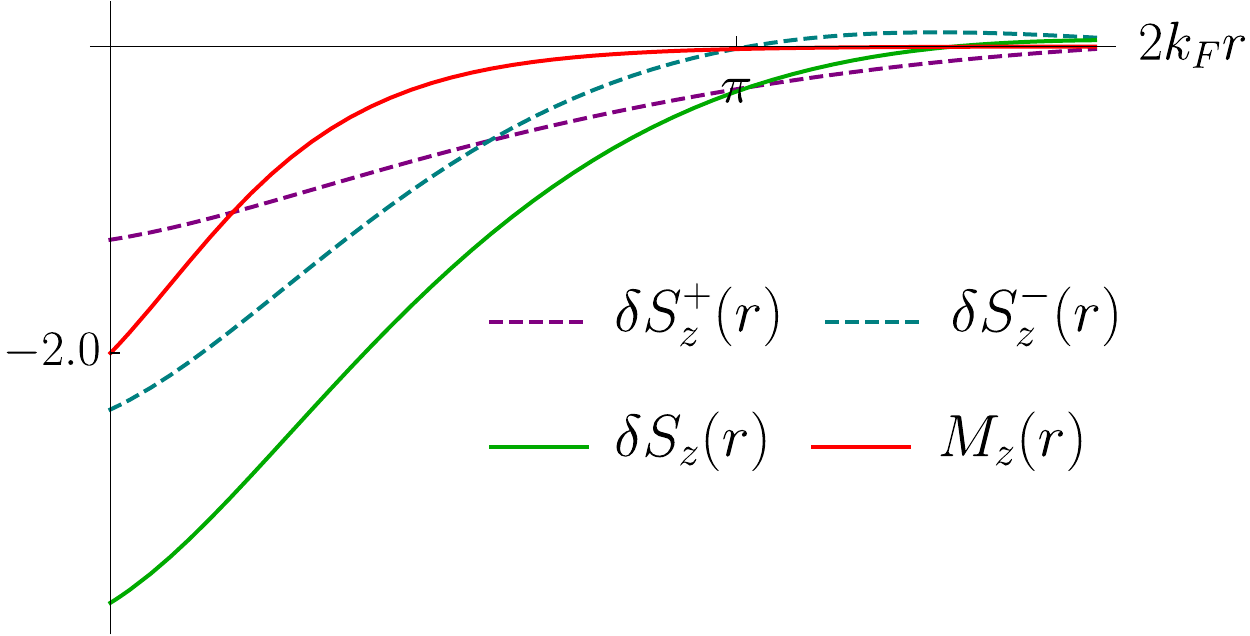}
    \caption{Spin density response $\delta S_{\parallel, z}$  induced by a small skyrmion: $k_F a = 0.5$. Two subbands are filled: $\alpha/E_F = 0.8$;}
    \label{fig:4}
\end{figure*}

The situation is different when $k_F a \lesssim 1$ (equivalent to $k_F^\pm a \lesssim 1$, as ${\alpha}/{E_F} < 1$).
This case is shown in Fig.~\ref{fig:4}.
The spin density response $\delta S_{\parallel,z} (r)$ has a
characteristic spatial scale of $\sim 1/k_F$, which is larger than the localization $a$ of the magnetization field $M_{\parallel, z}$. This non-local dependence has a clear physical explanation. 
When the electrons try to screen the impurity 
they can not come closer than their minimal wave length, which is $2 \pi/k_F$ (or $\sim 2 \pi/k_F^\pm$). Significant Friedel oscillations emerge and decay as $1/r^2$. 
As mentioned above $\delta S_{\parallel}$ oscillates with the spatial period $2\pi(q_*)^{-1}$ while $\delta S_{z}$ oscillates with periods two $\pi(k_F^+)^{-1}$ and $\pi(k_F^-)^{-1}$.

\begin{figure}[ht]
    \centering
    \includegraphics[scale = 0.7]{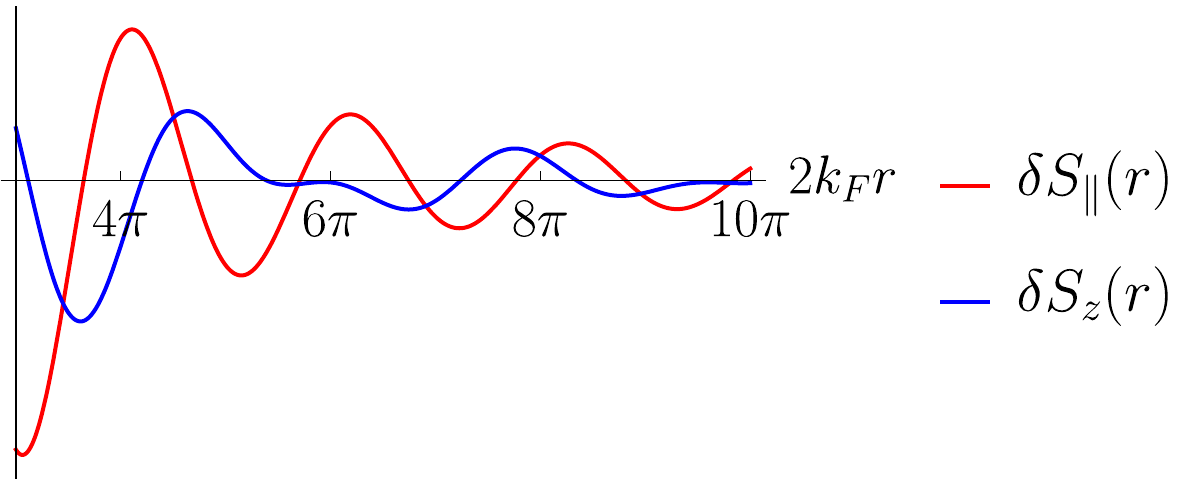}
    \caption{Phase difference of $\delta S_{\parallel}$ and $\delta S_{Z}$ oscillations. Small skyrmion: $k_F a = 0.5$. Two subbands are filled: $\alpha/E_F = 0.8$.}
    \label{fig:5}
\end{figure} 

As we have previously noted, the spin density response has the form of Eq.~\ref{eq:13}, which looks similar to that of the skyrmion magnetization field Eq.~\ref{eq:2}. 
However, 
there is an important difference between the amplitudes $\bar{M}_{\parallel, z}$ and $\delta S_{\parallel, z}$.
As our calculations show, a phase difference emerges between the $\delta S_{\parallel}$ and $\delta S_{z}$ Friedel oscillations (Fig.~\ref{fig:5}).
Therefore, a local spin density vector would 
rotate 
along any given direction when 
passing through the center of the skyrmion. 
So, the skyrmion induces a chiral spin structure in the electron gas, which appears to be 
different and more complex than the structure of the skyrmion itself. 

\subsection{One subband case}
\begin{figure}[ht]
    \centering
    \includegraphics[scale = 0.9]{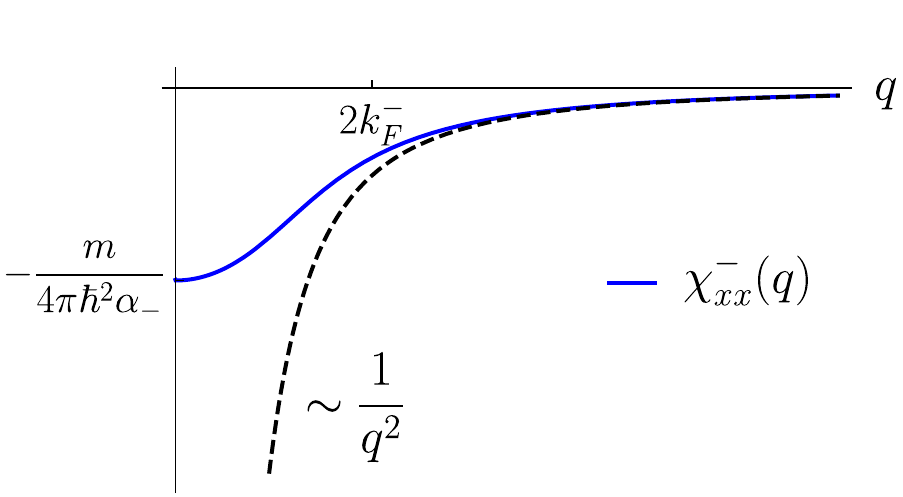}
    \caption{$\chi_{\parallel}$ spin-density response function at $\frac{\alpha}{E_F} = 10$ (only one subband is filled). At zero the function is quadratic, at infinity inverse quadratic.}
    \label{fig:6}
\end{figure}

\begin{figure*}[t]
    \centering
    \includegraphics[scale = 0.6]{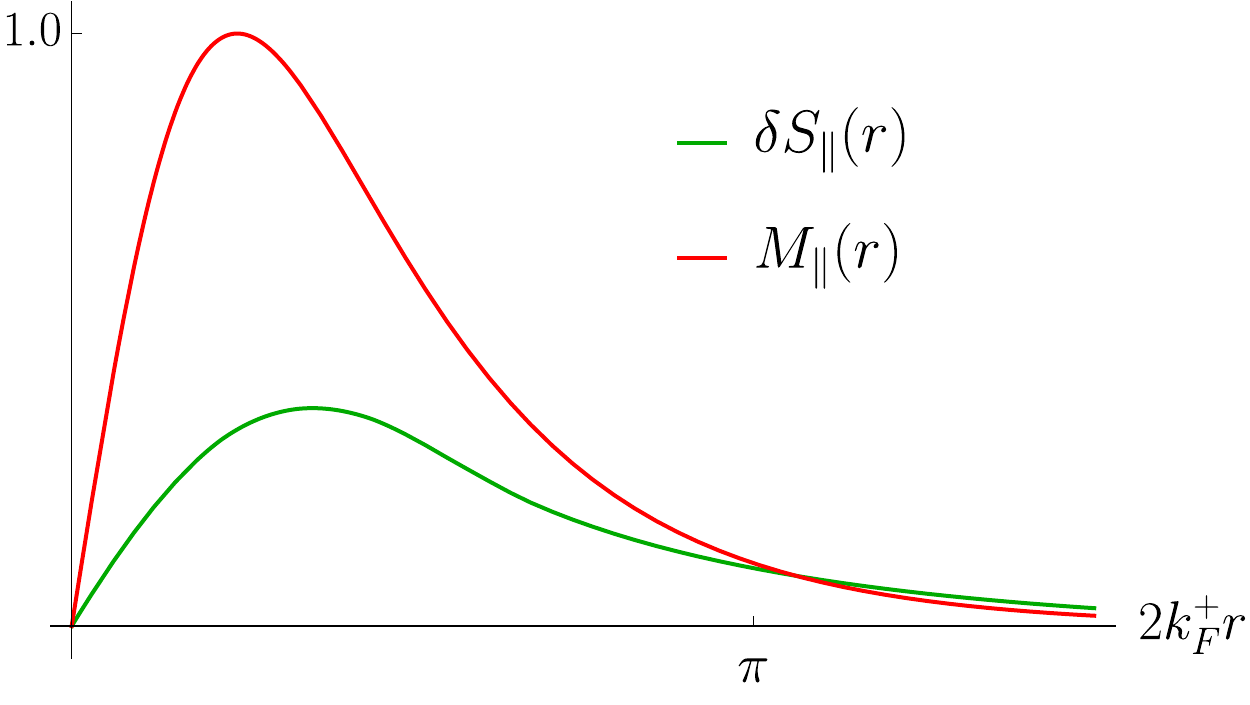}
    \includegraphics[scale = 0.6]{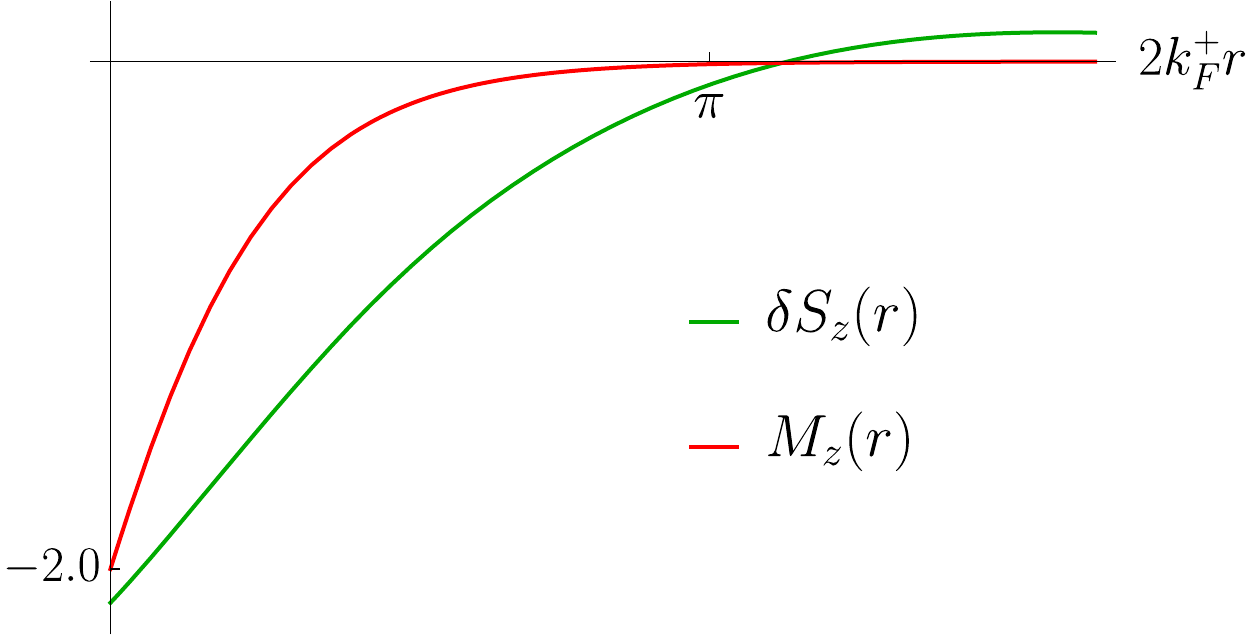}
    \caption{$\delta S_{\parallel, z}$ spin density response induced by a small skyrmion: $k_F a = 0.5$. Only one subband is filled: $\alpha/E_F = 2$.}
    \label{fig:7}
\end{figure*}

In the case $E_F \leq \alpha$ there is no contribution of the upper spin subband $(+)$.
Consequently, the function $\chi_z$ has only one
critical point, 
the corresponding Friedel oscillations are characterized by a single spatial period $\pi/k_F^{-}$.
The function $\chi_\parallel$ also has a different shape (see Fig.~\ref{fig:6}). Namely, the absence of critical points for 
$\chi_{\parallel}$ results in the absence of 
the oscillating structure of $\delta S_\parallel$ decaying as  $ \propto r^{-2}$. 
Instead, the amplitude of the oscillations having spatial period $2\pi/k_F^-$  decreases
exponentially (see Fig.~\ref{fig:7}). The magnitude of the $\delta S_\parallel$ response is proportional to ($\alpha + E_F$) and vanishes when $E_F$ approaches the bottom of the lower subband. This is due to the fact that only one subband is filled, so it is harder to mix both states to get the spin response in the $xy$ plane.
In the case $k_F a \gg 1$ the physical picture does not change much from the case of the two filled subbands, the local dependence between spin density response $\delta \boldsymbol{S}$ and $\boldsymbol{M}$ remains. 
When $k_F a \lesssim 1$ the spin density response has the effective size  $1/k_F$. 


 
\section{Summary and Discussion}

We have studied the spin structure of an electron gas induced by a magnetic skyrmion. We have shown that the skyrmion induces a chiral distribution of the electron gas spin density. This spin density spatial distribution turns out to be more complex than the skyrmion structure itself. There are two physically different regimes depending on the size of the skyrmion and the Fermi energy. At $k_F a \gg 1$ the electron spin response and the magnetization field of the skyrmion are locally coupled, that is the 'classical' regime.
The opposite case of  $k_F a \lesssim 1$ can be referred as 'quantum' regime as in this case 
the wave nature of the electrons comes to the fore.
The induced electron spin structure has now a minimal size of $1/k_F$ and pronounced Friedel oscillations. In the case of only one occupied spin subband   $\delta S_\parallel$ Friedel oscillations are suppressed.

Let us estimate the considered effect of skyrmions on the electron spin density 
for various systems. 
The stabilization of individual skyrmions up to the room temperature has been recently achieved in 
stack heterostructures consisting of atomically thin metal layers (Ir,Fe,Co) 
in contact with heavy metal elements, such as Pt~\cite{fert2017magnetic,raju2017evolution}.   
With a typical skyrmion size being $50$ nm~\cite{soumyanarayanan2017tunable} we expect an electron gas in any of the metallic layers 
to be \textit{locally} coupled with magnetization (the parameter $k_F a \gg 1$). 
On the other hand, an intriguing situation can be realized in so-called proximitized materials~\cite{Zutic2018} 
if 
a ferromagnetic layer hosting skyrmions is put 
in a contact with a semiconductor cap. 
Taking a typical Fermi wavelength for 2DEG in a semiconductor media as $k_F \sim 10^{6}$ cm$^{-1}$ and skyrmion size $a=10$ nm
we get $k_F a \approx 1$, which indicates that the non-local regime is realized. 
As a consequence, 
one would expect 
the emergence of RKKY-like 
interaction between two skyrmions in the same ferromagnetic layer 
due to the non-locality of 2DEG response in the proximitized semiconductor layer. 





\section*{Acknowledgments}
The work
has been carried out under the financial support
from Russian Science Foundation,
project 17-12-01182 (analytical theory), 
N.S.A., K.S.D., I.V.R, L.A.Y. thank the Foundation for the Advancement of Theoretical Physics and Mathematics "BASIS".
The work was also supported by  Russian Foundation of Basic Research (grant 18-02-00668) 
 and by the Academy of Finland (grant No.~318500).
\appendix

\section{Response functions}

\label{A1}

The general expression for the spin susceptibility is as follows
\begin{multline}\label{eq:8}
\chi_{\alpha \beta} (\boldsymbol{q})= \frac{1}{L^d} \sum\limits_{\boldsymbol{k},s,s'} \frac{f_{\boldsymbol{k}}^s - f_{\boldsymbol{k}+ \boldsymbol{q}}^{s'}}{\varepsilon_{\boldsymbol{k}}^s - \varepsilon_{\boldsymbol{k} +\boldsymbol{q}}^{s'} + i0} \times \\ \times \bra{u_{\boldsymbol{k}}^s} \hat{S}_\alpha \ket{u_{\boldsymbol{k} + \boldsymbol{q}}^{s'}} \bra{u_{\boldsymbol{k} + \boldsymbol{q}}^{s'}} \hat{S}_\beta \ket{u_{\boldsymbol{k}}^s},
\end{multline}
where $d = 2$ is the dimension of our system, $L$ is the normalization length,
 $s$ and $s'$ are spin indexes, $\hat{S}$ is the electron spin operator, $f_{\boldsymbol{k}}^s$ is the electron distribution function; $\varepsilon_{\boldsymbol{k}}^s$ and $u_{\boldsymbol{k}}^s$ are the energy 
 and the spin state, respectively. The mentioned above variables relate to the unperturbed system with Hamiltonian $\hat{H}_0$. 
 Given the system energy spectrum and the distribution function 
 the spin-density response function $\hat{\chi}$ can be obtained. 
We assume zero temperature, so $f_{\boldsymbol{k}}^\pm =  \theta\left(k_{F}^\pm - k\right).$

\section{Structure of $\chi_\parallel$}
\label{A2}
The expression for $\chi_{xx}$ with contributions from both spin subbands appears to be 
\begin{multline}
\chi_{xx} = - \frac{m}{2 \hbar^2 \pi} \frac{1}{z}\left[2z - \theta\left(z^2 +\frac{\beta^2}{z^2} - 1\right) \times\right.\\\left.\times \sqrt{z^2 +\frac{\beta^2}{z^2} - 1} (\sgn(z - \beta/z) + 1) \right],
\end{multline}
where $z = \frac{q}{2 k_F}$ and $\beta = \frac{\alpha}{2 E_F} < \frac{1}{2}$, 
as the equation remains valid when both subbands are occupied. 
The $q_*$ critical point appear when the second term in the above equation becomes non-zero. The second term is non-zero when both $\sgn(..) = 1$ and $\,\theta(..) = 1$. Solving the quadratic equation in the argument of Heaviside $\theta$ function we get a solution $z_*$, which satisfies the equation in the Sign function argument, i.e. $z_* - \beta/z_* > 0$, assuming $\beta <\frac{1}{2}$.

\begin{gather}
z_*^2 + \frac{\beta^2}{z_*^2} - 1 = 0 \Rightarrow z_* = \sqrt{\frac{1 + \sqrt{1- 4 \beta^2}}{2}} > \sqrt{\beta},\\
q_* = 2k_F \cdot \sqrt{\frac{1 + \sqrt{1- 4 \beta^2}}{2}},\,\,\,\,\,
2k_F^+ \leq q_* \leq  2k_F^-.
\end{gather}

\bibliography{Skyrmion}

\end{document}